# On the roughness of single- and bi-layer graphene membranes

J. C. Meyer (1), A. K. Geim (2), M. I. Katsnelson (3), K. S. Novoselov (2),

D. Obergfell (4), S. Roth (4), Ç. Girit (1), A. Zettl (1)

(1) Department of Physics, University of California at Berkeley, and Materials Sciences Division, Lawrence Berkeley National Laboratory, Berkeley, California 94720 U.S.A.
(2) Manchester Centre for Mesoscience and Nanotechnology, University of Manchester, Oxford Road, Manchester M13 9PL, United Kingdom
(3) Institute for Molecules and Materials, Radboud University of Nijmegen, Toernooiveld 1, 6525 ED Nijmegen, The Netherlands
(4) Max Planck Institute for Solid State Research, Heisenbergstr. 1, 70569 Stuttgart, Germany

PACS: 81.05.Uw, 82.45.Mp, 61.14.Lj, 61.48.+c, 68.55.-a, 46.70.Hg

Keywords: Graphene, membranes, transmission electron microscopy

## Abstract

We present a detailed transmission electron microscopy and electron diffraction study of the thinnest possible membrane, a single layer of carbon atoms suspended in vacuum and attached only at its edges. Membranes consisting of two graphene layers are also reported. We find that the membranes exhibit an apparently random spontaneous curvature that is strongest in single-layer membranes. A direct visualization of the roughness is presented for two-layer membranes where we used the variation of diffracted intensities with the local orientation of the membrane.

The recent discovery of graphene in a quasi-free state [1, 2] has sparked considerable interest in the science and applications of this new material with its remarkable electronic properties. Graphene monolayers on semiconductor substrates can be patterned and contacted by conventional lithographic techniques, and the resulting devices provide access to the rich physics of quantum electrodynamics in a solid state physics experiment [3, 4]. However, graphene provides not only a two-dimensional electronic system. It also provides the best possible approximation to a two-dimensional material in all other aspects, e.g. with respect to lattice vibrations [5-7] or as a mechanical system [8].

Indeed, the apparently 2D structure itself is one of the most intriguing properties of graphene. In a graphite crystal, the atoms within individual planes are arranged on a strictly 2D honeycomb lattice with strong in-plane (sp2) bonds and only a weak (van der Waals-like) out-of-plane interaction. Even after the separation of individual planes by mechanical cleavage, the graphene sheets sustain a sufficiently ordered state to support sub-micron mean free paths. This is of particular significance in the view of theoretical arguments which show that 2D crystals should not exist [9-14]. However, in most previous experiments, graphene layers were supported by a substrate, or embedded in a bulk material. We have prepared freely suspended membranes of singe- and few-layer graphene and analyze their structure by transmission electron microscopy (TEM). Our analysis shows that these two-dimensional membranes do not remain flat but show a roughness that appears to be intrinsic to graphene membranes.

To make graphene membranes, we started with the established procedure of micromechanical cleavage [1-4] and obtained graphene crystallites on top of a silicon substrate with a 300nm silicon dioxide layer. Potential mono-layer sheets were identified by optical microscopy or scanning electron microscopy, and located with respect to a marker system. A metal grid with 400nm-wide bars separated by distances between 400nm and 1000nm was then deposited on the graphene sheets by electron beam lithography and thermal evaporation of 3nm Cr and 100nm Au. The substrate was cleaved through the grid, close to the graphene so that it is within ≈ 50µm of the cleaved edge. We then etched the bulk silicon from the side of the cleaved edge, leaving the metal grid with the graphene membrane extending over the edge of the substrate. The first etching step consisted of several hours in 15% tetramethylammonium hydroxide at 60°C, which removed the bulk silicon but left the oxide layer and the metal grid in place. This etching step was monitored with an optical microscope and stopped as soon as a sufficient part of the metal grid was undercut. Next, the free-standing part of the oxide layer was removed by 5 minutes in 6% buffered hydrofluoric acid. Finally, a critical point drying step was required to preserve these delicate structures, so that the graphene membranes remained attached to the grid. The free-standing part of the grid then became accessible by TEM. A similar process was used previously to obtain fragile TEM-compatible carbon nanotube devices [15,16].

Figure 1 shows a suspended graphene membrane with a lateral size of several micrometers. This sample was identified as two layers by electron diffraction. In the bright-field TEM images, these membranes provided no detectable absorption and were only visible as phase contrast at a sufficiently large defocus. Folds and scrolls are seen at the rim of the membrane (Fig. 1 inset), whereas large parts of the inner regions appears featureless. Occasionally, adsorbates of unknown origin were found, such as those visible in Fig. 1, but they tend to cluster and leave most of the membrane clean.

Figure 2a shows an example of a single layer graphene membrane. The regions indicated by the arrows are an individual layer of carbon atoms suspended in vacuum and attached only at its edges, as verified by electron diffraction. In this dark-field TEM image the incident beam and objective aperture have been set to select only electrons that were scattered by a small angle, by tilting the primary beam to just outside the aperture. Since no Bragg reflections are selected in this way, an image of the thickness is obtained for a single element sample. Accordingly, the gray levels in the folded areas are integer multiples of that in the single-layer area. This imaging mode is also very sensitive to surface adsorbates [17] and the homogeneous appearance in Fig. 2 indicates a good purity of the membrane.

The number of layers in a suspended graphene membrane can be determined by nanoarea electron diffraction patterns by varying incidence angles between the electron beam and the graphene sheet. This approach effectively probes the whole 3D reciprocal space. Fig. 3a,b shows calculated 3D Fourier transforms of single- and bi-layer (AB stacked) graphene atom positions (Note that the atomic scattering factors are not incorporated here, so that the intensities in Fig. 3a,b are only qualitatively correct). High intensity volume elements in this 3D data set are visualized by an isosurface, and a section through the data corresponding to the normal incidence diffraction pattern is shown as colours on the blue plane. For a single layer planar crystal the reciprocal space is a set of rods (arranged on the 2D reciprocal lattice) with a weak, monotonous intensity variation normal to the plane. The intensity profile along any of the rods is given only by the product of the atomic form factor and the effective Debye-Waller factor. For two (or more) layers, an additional modulation appears. As a consequence, variations of a few degrees in the tilt angle lead to strong variations in the diffraction intensities for all multi-layer samples independent of the stacking sequence, which allows a direct and unambiguous identification of single- vs. multi-layer samples.

Fig. 3(e-h) shows the variation of diffraction intensities with tilt angle for a single- and bi-layer membrane, both as experimental data (solid lines) and numerical simulations (dashed lines). These simulations are obtained by a Fourier transform of projected atomic potentials, and are based on the scattering factors of Ref. [18]. The simulation takes into account that the number of atoms within a beam increases as the sample is tilted, leading to a slight increase in intensity with tilt angle in some peaks. We use the Bravais-Miller (hkil) indices to label the peaks equivalent to the graphite reflections at normal incidence, although, strictly speaking, the index would be different for the bi-layer reflections after tilting through a minimum. It was found that all our few-layer samples, prepared by mechanical cleavage of graphite, retain the Bernal (AB) stacking of the source material. Once this is established, the single layer membranes can be identified from a normal incidence pattern only, by analyzing the intensity ratio of the Bragg reflections (Fig. 1). However, we note that AA... stacking has been reported in carbon nanofilms produced by another technique [19].

The tilted incidence patterns provide insight into structural modifications that occur in these atomically thin membranes: namely, they reveal deviations from the idealized graphene sheet. While the total (integrated) intensities within each Bragg reflection agree quite well with the model of a flat membrane, the actual shape and widths of the peaks show striking deviations from the standard diffraction behavior of 3D crystals [20,21]. Fig. 5 shows two diffraction patterns obtained from a single-layer membrane at incidence angles of 21° (a) and 28° (b). We observe that the peaks broaden with increasing incidence angle, such that a sharp peak at normal incidence spreads isotropically to an approximately Gaussian shaped smooth intensity distribution. Moreover, the widths of the Gaussian fits are roughly proportional to the tilt angle and to the distance of the peak from the tilt axis (that is, peaks close to the tilt axis remain rather sharp). This effect is very prominent in single-layer membranes but it is significantly reduced in two-layer samples and not present in thin graphite. For comparison, Fig. 5(c) shows the diffraction pattern from a two-layer membrane under the same conditions as for the monolayer in Fig. 5(a). In fact, Figs. 5 (a-c) were obtained from the same sample, with the membrane being single-layer in one half of the area and bi-layer in the other half. Therefore, Fig. 5 (a) and (c) were obtained at precisely the same angle, orientation and imaging conditions, with the only difference being the number of layers.

The peak broadening can be understood by assuming that the graphene membrane is not exactly flat. If we model the membrane as a number of locally flat pieces with slightly different orientations, each piece yields a diffraction peak at a slightly different position, and their incoherent superposition leads to diffraction intensities that do not fall onto a single point. In the reciprocal space, this can be understood as a superposition of rods with slightly different orientations (Fig. 6), so that the diffraction intensities are different from zero in a cone-shaped volume. This model also predicts that the peaks are sharp at normal incidence and their width increases linearly with tilt angle, as indeed observed experimentally. Fig. 6f shows the FWHM width of diffraction peaks with tilt angle. The linear slope can be directly related to the cone angle, which is found to be ≈10° in monolayer samples. This means the surface normals deviate by ca. 5° from the mean surface. For bi-layer membranes, the spot broadening is approximately half as strong as in monolayers, which implies a mean deviation of ≈2°.

Although in bi-layer samples the curvature of the membrane is only approximately half as strong as compared to monolayers, it is easier to visualize directly. We could observe the roughness of bi-layer membranes by using the strong variation in their diffraction intensities with tilt angle (Fig. 4g,h). We used convergent beam electron diffraction (CBED) with the sample offset from the beam focus to map out the diffraction intensities for a portion of the sample (Fig. 7). In this way, each diffraction spot allowed us to image the illuminated area for this particular Bragg reflection [20]. Since these intensities are obtained through Bragg reflections from the crystal lattice, they depend only on the local orientation of the graphene membrane and cannot be an image of adsorbates or defects. The variation of the incident beam angle across the sample due to the convergent probe can be neglected for a two-layer graphene membrane[1]. The resolution is approximately given by the size of the spot at the beam focus in Fig. 7d. Reducing the spot size leads to strongly decreased diffracted intensities as compared to the undiffracted beam, which effectively limits the resolution (this can be seen by comparing Fig. 7f, that exhibits a good signal-to-noise ratio but shows only large scale variations, with Figs. 7g-j, in which the spot size is optimized to obtain a better resolution at the expense of a noisier signal). The variation in local orientation of the membrane leads to intensity variations within the CBED spots which is in ageement with the ca. ±2° deviation from the average normal that was inferred from the broadening of the spots in the nanoarea diffraction patterns. Figs. 7g-j show ripples with a lateral extent down to 15nm, which we estimate as the resolution limit in this configuration. Ripples of smaller lateral extent (few nm) have been observed by atomic resolution TEM imaging [22]. Importantly, the ripples are found static (CBED patterns are reproduced at subsequent exposures, see Fig. 7) and have an apparently random distribution of lateral sizes, orientation and heights. We note again that the grey scales in Fig. 7g-j correspond to different orientations rather than heights: One can view the image in Fig. 7h as a curved landscape illuminated at a grazing angle from the lower left corner, like mountains at sunrise. The actual shapes of the ripples becomes clear in this way; and this appearance is well justified by the underlying contrast mechanism.

Our TEM studies show that the suspended graphene membranes assume a static, non-flat configuration with apparently random microscopic out-of-plane deformations where surface normal varies by several degrees. The smooth Gaussian shaped broading implies that there are large number of different orientations present within the submicron diameter electron beam, and that the surface normal must vary in all directions. This means that there is a microscopic roughness present within our membranes with no preferred orientation. The reproducible appearance across samples indicates that it is an intrinsic effect. It is important to note that the homogeneous and isotropic broadening we observe is not compatible with bending deformation of a rigid membrane. This contradicts the assumption of an incompressible sheet, which could be curved in one but not two directions. To emphasize this point, we remind that, for example, a sheet of paper (which has a very high in-plane elastic modulus) can be curved into a cylinder but not into (a section of) a sphere. Consequently, the observed broadening can not be explained by strain-free deformations of graphene. We estimate local strains of up to 1% for the single-layer membranes.

From a theoretical point of view, graphene is an example of a crystalline membrane, or can also be described as a polymerized membrane or a tethered membrane. This class of membranes is predicted to exist in three different configurations: a flat one, a so-called crumpled phase with a fractal dimension, and a compact (collapsed) phase [23]. While most calculations assume a free membrane, our graphene sheets are attached to a solid frame that provides a boundary condition. This forces the membrane to be flat on the spatial scale of the supporting frame. In this study we have focused on these nearly flat regions, and observed static ripples. Note that, strictly speaking, ideal 2D crystals are thermodynamically unstable at a finite temperature [24-26]. The apparent stability of graphene membranes can be due to the fact that they are quenched in a meta-stable configuration after being extracted from 3D (that is, stable) graphite at a relatively low temperature so that strong interatomic bonds and small sample sizes do not allow the generation and propagation of crystal defects. An interesting alternative is that the observed rippled configuration can increase the thermodynamic stability of graphene membranes, and may in fact be energetically more favorable [24-26]. While graphene on a substrate is supported on the entire area, the membranes have a higher degree of freedom that would allow them to move towards a more favored configuration as far as permitted by the supporting frame. In fact, we also observed crumpled sheets that became partly detached from the metal grid (Fig. 8), and we expect that further studies of these structures, as well as of scrolled and folded regions near membrane edges, will provide further insights into the interesting problem of stability of 2D crystal systems. In any case, more experimental and theoretical work is required to understand the observed roughness.

---

[1] This is in contrast to CBED experiments on bulk crystals, where often the whole point of the CBED pattern is to map out an intensity vs. incidence angle dependence. However, for only two layers, significant intensity variations require tilting by a few degrees (Fig. 3g,h) instead of tiny fractions of a degree as in bulk crystals.

In conclusion, single-layer graphene can be used to make the thinnest possible membranes, with a thickness of just one atom. We present an unambiguous identification of single- and bi-layer samples by nanoarea electron diffraction. The membranes are not flat but exhibit random out of plane deformations. The one atom thick membranes are intriguing objects for research that are strikingly different from ordinary three-dimensional crystals.

Acknowledments: This work was supported in part by the Director, Office of Energy Research, Office of Basic Energy Sciences, Materials Sciences Division of the US Department of Energy under Contract No. DE-AC-03-76SF00098 and by the National Science Foundation under Grant N. EEC-0435914, supporting the Center of Integrated Nanomechanical Systems, the EU project CANAPE, the EPSRC (UK) and the Royal Society. M.I.K. acknowledges financial support from FOM (Netherlands).

References
[1] Novoselov, K. S. et al. Electric field effect in atomically thin carbon films. Science 306 (2004) 666-669.
[2] Novoselov, K. S. et al. Two-dimensional atomic crystals. Proc. Natl Acad. Sci. USA 102 (2005) 10451-10453.
[3] Novoselov, K. S. et al. Two-dimensional gas of massless Dirac fermions in graphene. Nature 438 (2005) 197-200.
[4] Zhang, Y., Tan, J.W., Stormer, H.L. and Kim, P. Experimental observation of the quantum Hall effect and Berry's phase in graphene. Nature 438 (2005) 201-204.
[5] A. C. Ferrari et al., The Raman Spectrum of Graphene and Graphene Layers, Phys. Rev. Lett. 97 (2006) 187401.
[6] A. Gupta et al., Raman Scattering from High-Frequency Phonons in Supported n-Graphene Layer Films, Nano Lett. 6 (2006) 2667-2673.
[7] D. Graf et al., Spatially Resolved Raman Spectroscopy of Single- and Few-Layer Graphene, Nano Lett. 7 (2007) 238-242.
[8] J. S. Bunch et al., Electromechanical Resonators from Graphene Sheets, Science 315 (2007) 490-493.
[9] Peierls, R. E. Bemerkungen über Umwandlungstemperaturen. Helv. Phys. Acta 7 (1934) 81-83.
[10] Peierls, R. E. Quelques proprietes typiques des corpses solides. Ann. I. H. Poincare 5 (1935) 177-222.
[11] Landau, L. D. Zur Theorie der Phasenumwandlungen II. Phys. Z. Sowjetunion, 11 (1937) 26-3 .
[12] Landau, L. D. and Lifshitz, E. M. Statistical Physics, Part I. Pergamon Press, Oxford, 1980.
[13] Mermin, N. D. and Wagner, H. Absence of ferromagnetism or antiferromagnetism in one- or two-dimensional isotropic Heisenberg models. Phys. Rev. Lett. 17 (1966) 1133-1136.
[14] Mermin, N. D. Crystalline order in two dimensions. Phys. Rev. 176 (1968) 250-254.
[15] J. C. Meyer et al., Single Molecule Torsional Pendulum, Science 309 (2005) 1539-1541.
[16] J. C. Meyer et al., Electron diffraction analysis of individual single-walled carbon nanotubes, Ultramicroscopy 106 (2006) 176-190.

[17] D. K. Bhattacharya et al., Jpn. J. Appl. Phys. 36 (1997) 2918-2921.
[18] Doyle, P. A. and Turner, P. S. Relativistic Hartree-Fock x-ray and electron scattering factors. Acta Cryst. A 24 (1968) 390-397.
[19] Horiuchi, S. et al. Carbon nanofilm with a new structure and property. Jpn. J. Appl. Phys. 42 (2003) L1073-1076.
[20] Buseck, P. R., Cowley, J. M. and Eyring, L. High-Resolution Transmission Electron Microscopy. Oxford University Press, 1988.
[21] Spence, J. C. H. High-Resolution Electron Microscopy. Oxford University Press, 2003.
[22] J. C. Meyer et al., The structure of suspended graphene sheets, Nature 446 (2007) 60-63.
[23] G. Gompper and D. M. Kroll, Network models of fluid, hexatic and polymerized membranes, J. Phys. Cond. Mat. 9 (1997) 8795-8834.
[24] Nelson, D. R. and Peliti, L. Fluctuations in membranes with crystalline and hexatic order. J. Physique 48 (1987) 1085-1092.
[25] Radzihovsky, L. and Le Doussal, P. Self-consistent theory of polymerized membranes. Phys. Rev. Lett. 69 (1992) 1209-1212.
[26] Nelson, D. R., Piran, T. and Weinberg, S. Statistical Mechanics of Membranes and Surfaces. World Scientific, Singapore, 2004.

Figures

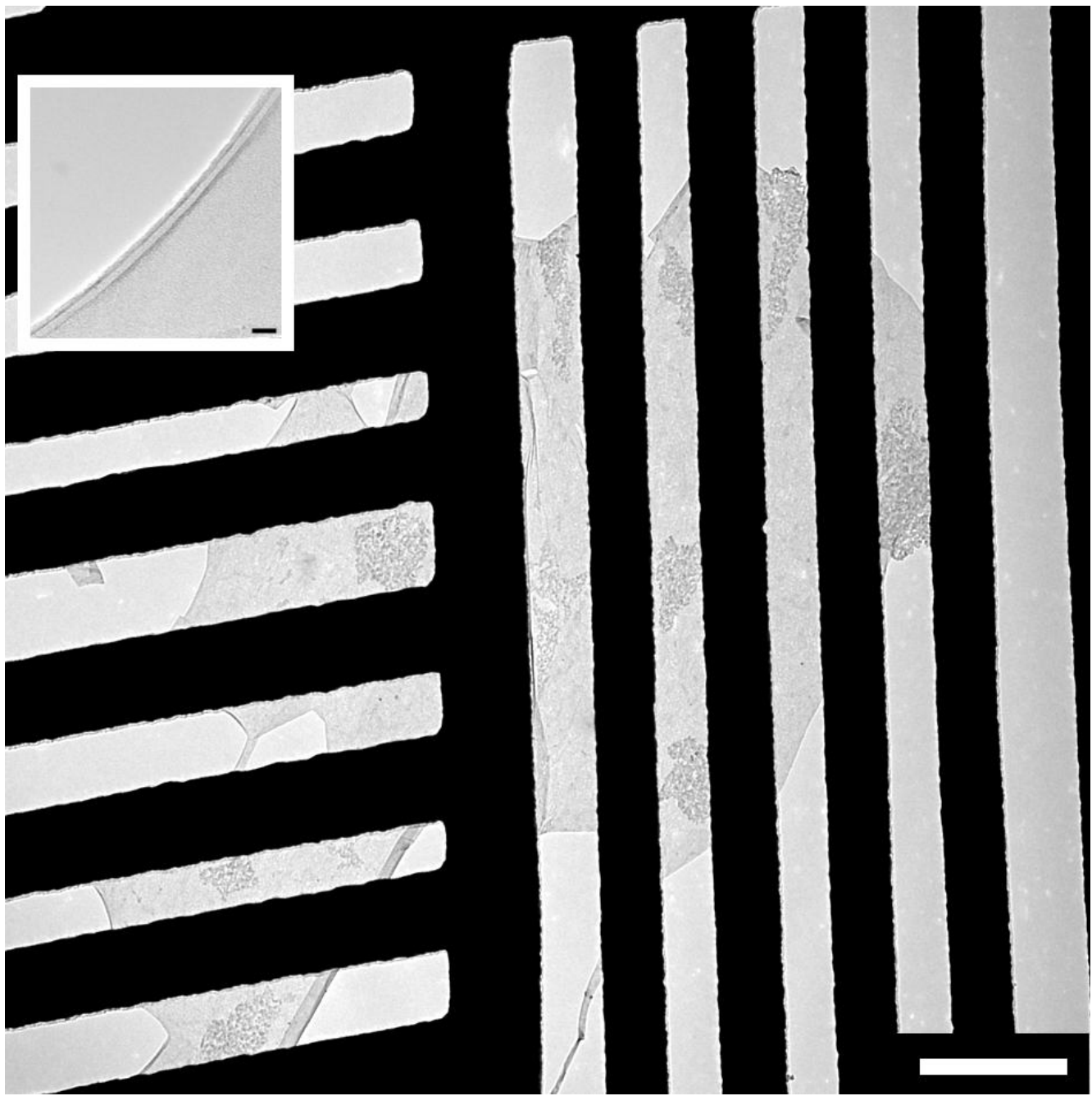

Figure 1: TEM image of suspended graphene (darker gray areas) supported by a microfabricated metal grid (black lines). The inset shows a scroll at the edge. Scale bar 1 μm, and 20 nm for the inset.

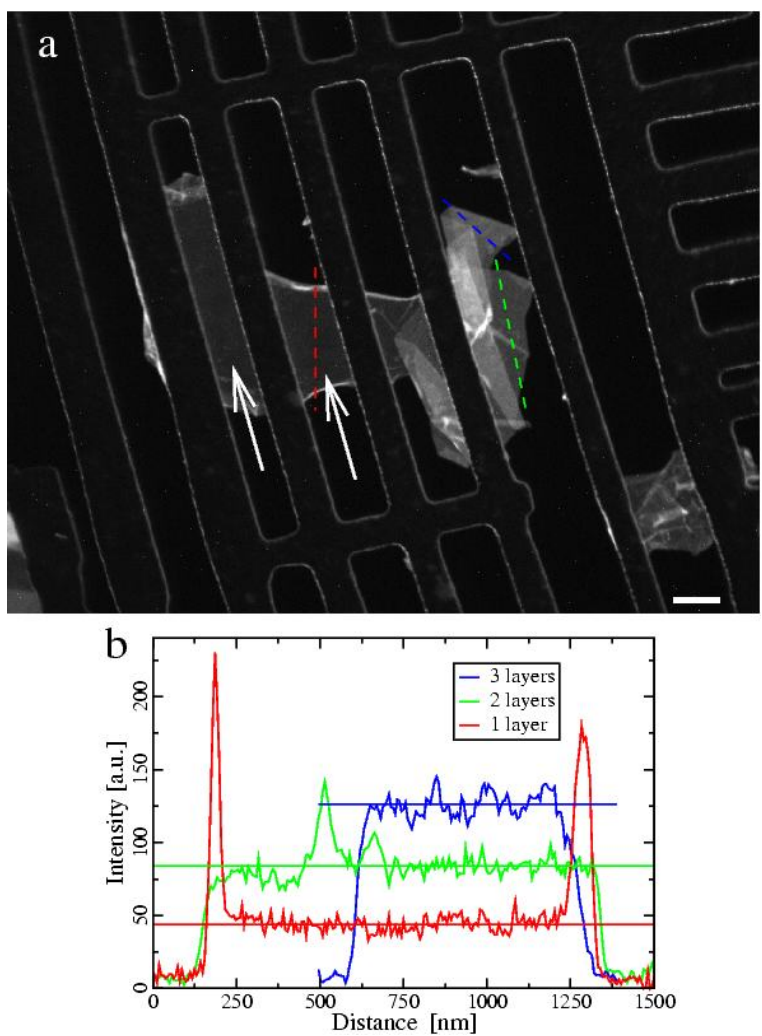

Figure 2: (a) Small angle dark-field TEM image of a single-layer graphene membrane. The dark grey area indicated by the white arrows is the single-layer region, as proven by electron diffraction measurements. Under these imaging conditions the intensity is proportional to sample thickness. The right part of the flake is folded, and indeed the recorded intensities in the folded areas are precisely integer multiples of the intensity in the single-layer area. Panel (b) quantifies this behavior by showing line profiles indicated with the respective colour in (a). In particular, the red profile plot exemplifies the very homogeneous appearance of the single-layer region, while green and blue lines show a double and triple folded region of the membrane. The horizontal lines are a guide to the eye. Scale bar in (a) is 500 nm.

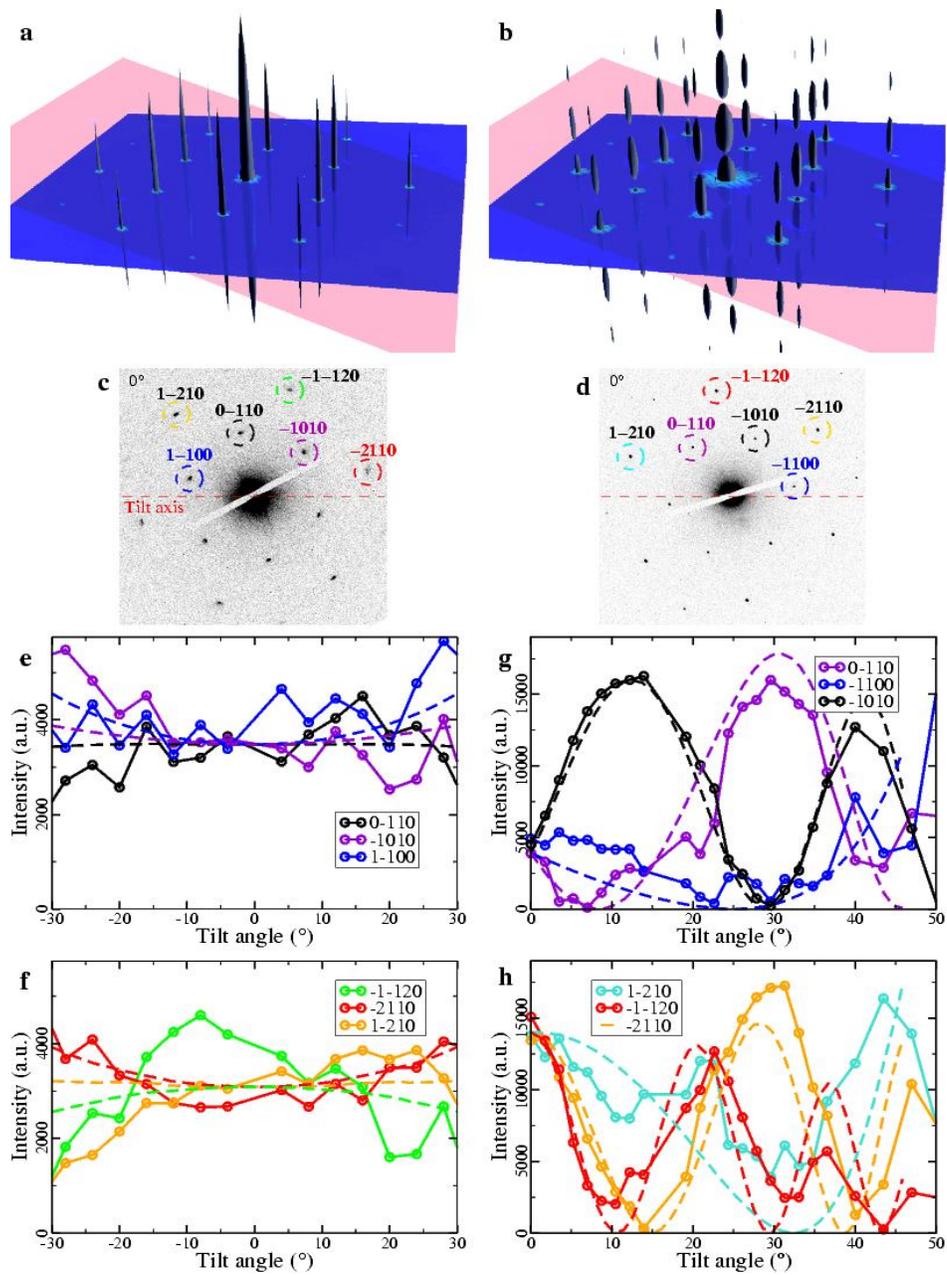

Figure 3: Reciprocal space of single- and double-layer graphene, as probed by electron diffraction experiments. (a) For single-layer graphene, intensities in reciprocal space are continuous rods so that the intensities in a diffraction pattern vary only weakly (only due to the atomic form factor) with the tilt angle between the membrane and the incident beam. (b) For two-layer samples, an intensity variation along the rods is present, and the diffraction peaks are suppressed at certain angles. The blue planes indicate a diffraction pattern that would be obtained at normal incidence, and the red plane for tilt angle of 20°. (c), (d) Normal incidence diffraction patterns of a single- and double-layer graphene membrane, respectively. The reflections plotted in (e-h) are indicated by the same colour. (e-h) Experimental data (solid lines) and electron diffraction simulations (dashed lines). (e,f) Intensities of diffraction peaks in the single layer membrane for a wide range of tilt angles (the plots are separated into two diagrams for clarity). The weak and monotonic variation is an unambiguous proof for a suspended monolayer. (g,h) The same analysis for the bi-layer membrane (d), showing the clear variations in the peak intensities with tilt angle. The behaviour matches simulations only for an AB stacked bilayer.

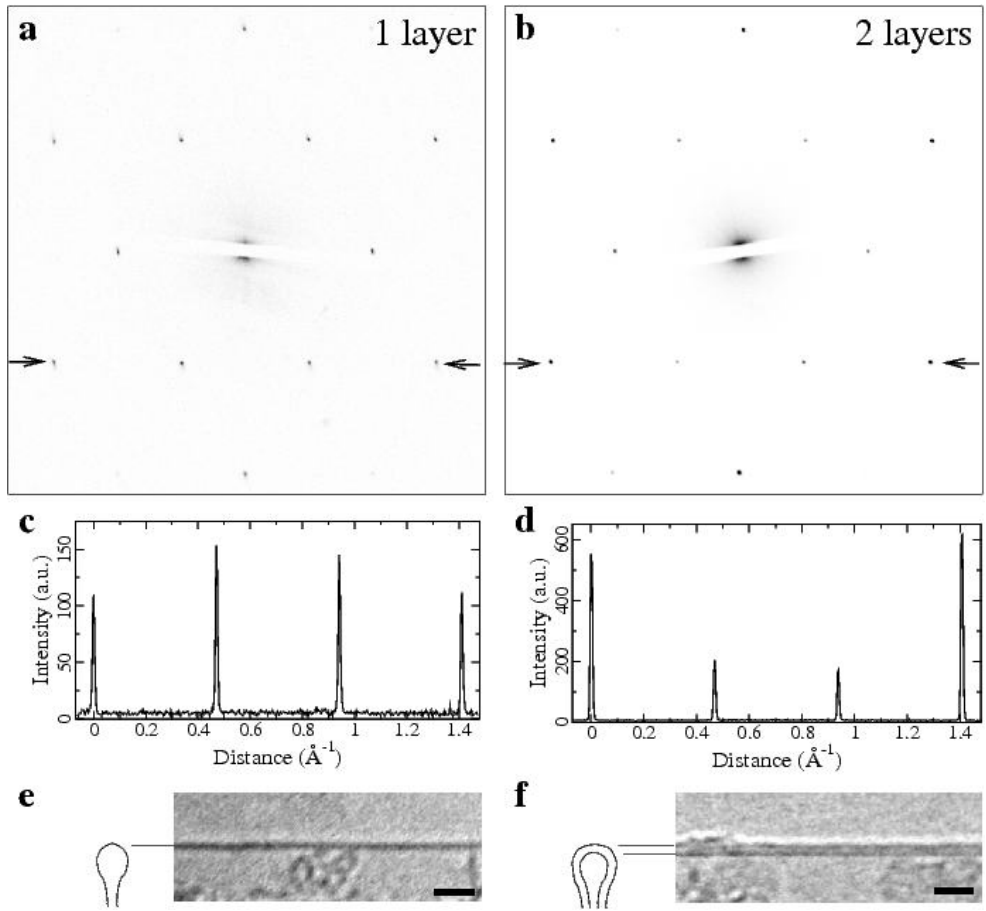

Figure 4: Nanoarea electron diffraction pattern of a single layer graphene membrane (a), and a two-layer membrane (b), at normal incidence. A profile plot along the line between the arrows is shown below in (c,d). If we assume that our samples always retain the Bernal (AB) stacking of the source graphite, the monolayer membranes can be identified already from the intensity ratios of the diffraction peaks (a definite identification for the number of layers and stacking sequence is obtained by the tilt series as shown in Fig. 3.). (e,f) Foldings at the rim of the membrane, where the sheet is locally parallel to the beam, show predominantly one line for single-layer samples (e), and two dark lines for two-layer samples (f) (see also Fig. 8d). Scale bars are 2 nm.

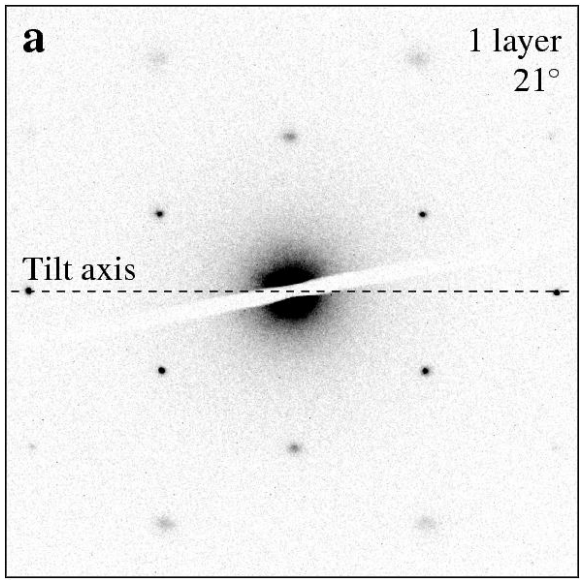

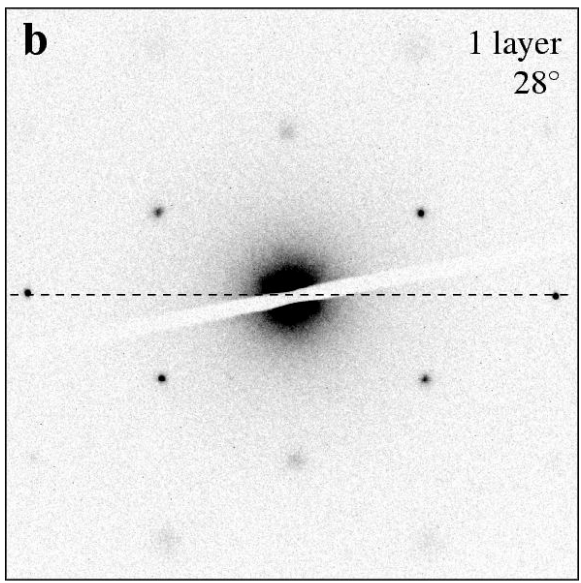

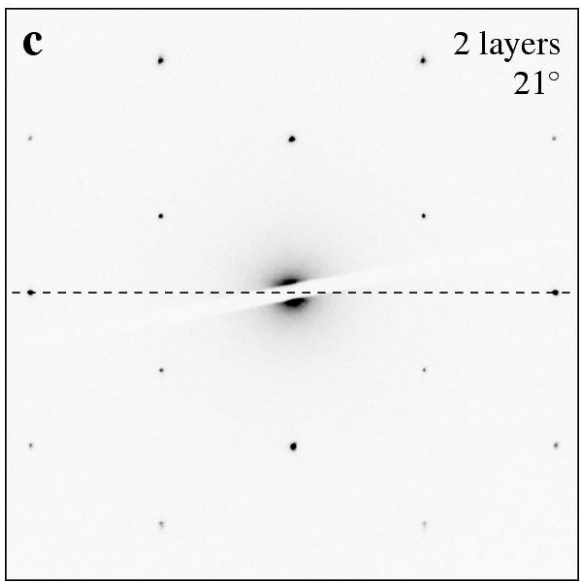

Figure 5: Tilted incidence diffraction patterns. (a), (b) Single layer graphene membrane at 21° and 28°, respectively. The peaks spread out into a smooth gaussian shaped intensity distribution with increasing tilt angle. (c) Two-layer graphene at 21° for comparison.

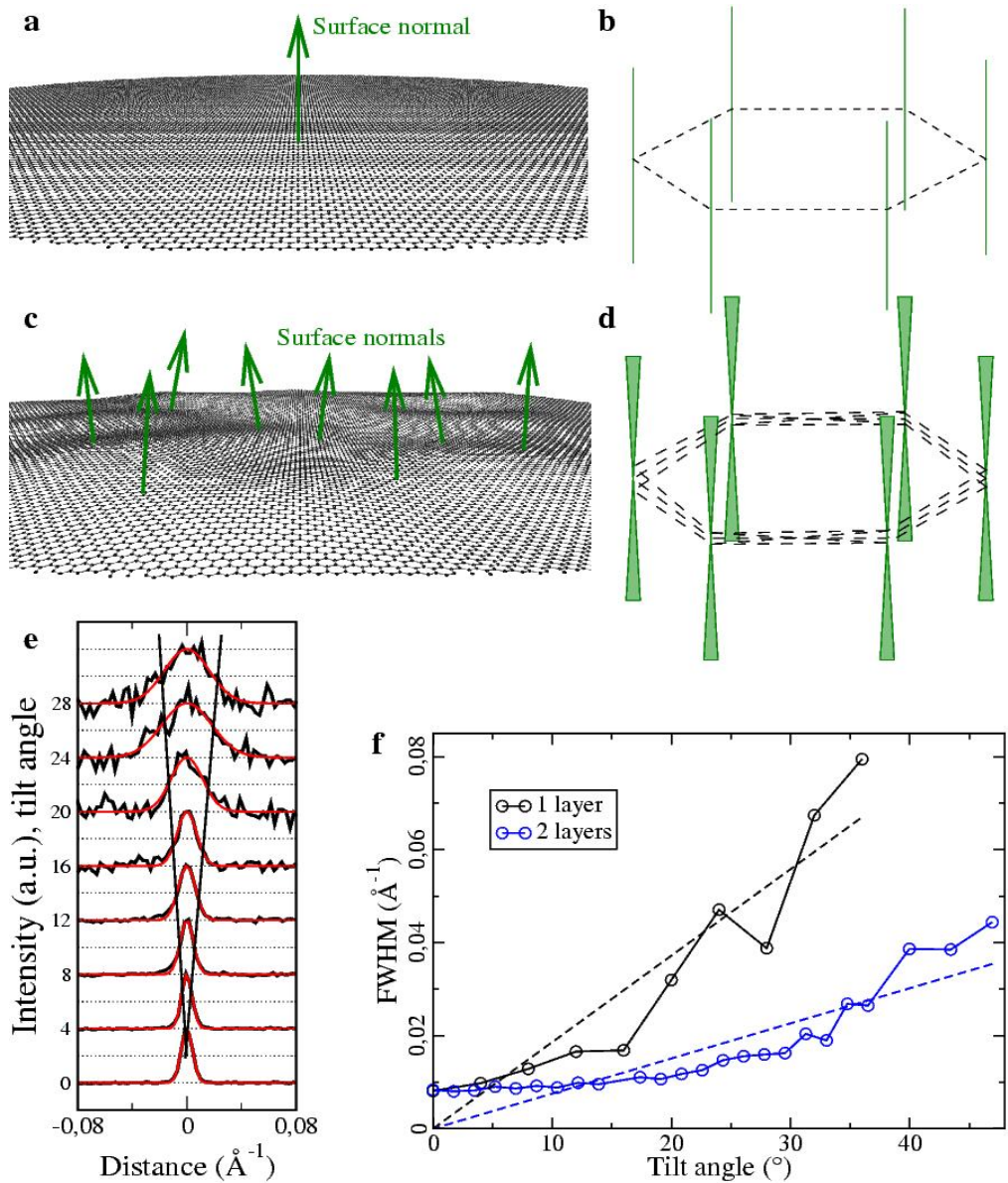

Figure 6: Explanation of peak broadening. For a flat graphene crystal (a), diffraction intensities constitute sharp rods in reciprocal space (b) that are parallel to the surface normal (also compare with Fig. 3a). If the surface is uneven (c), the diffracted intensities are obtained by a superposition of many rods with slightly different orientation (d). This gives rise to non-zero intensities in cone-shaped volumes in reciprocal space, and therefore to broadened diffraction peaks in the tilted incidence diffraction patterns. (e) Peak profiles (for the (0-110) reflection of Fig. 3c) for different incidence angles (black curves) and Gaussian fits (red), with an offset that corresponds to the tilt angle in degrees. The peak heights are scaled to the same size. A cone that connects the curves at approximately their FWHM is drawn as a guide to the eye. (f) FWHM of Gaussian fits for single- and bi-layer graphene vs. tilt angle. The slope (dashed lines) is proportional to the cone angle in (d). The peak broadening in bi-layer samples is approximately half as strong as in monolayers.

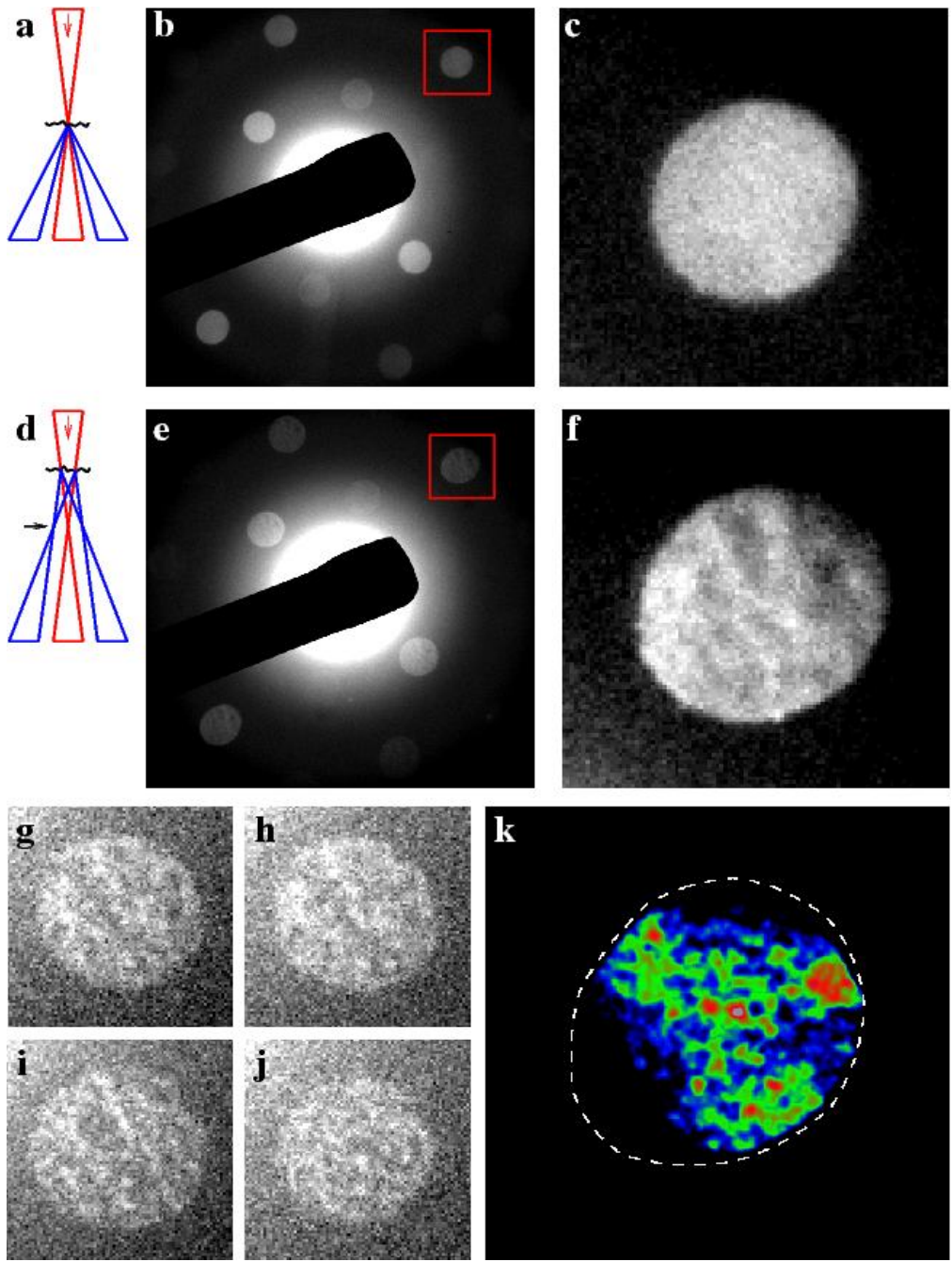

Figure 7: Convergent-beam electron diffraction (CBED) patterns from a two-layer graphene membrane. With the probe beam focused on the membrane (a), a pattern with smooth intensity discs (b,c) is obtained. If the sample is above (or below) the focus of the beam (d), each diffraction spot provides a mapping of the diffracted intensity for the illuminated area (e,f). Variations in the local orientation of the membrane translate into intensity variations inside the diffraction spots. The resolution is limited by the spot width at the crossover (indicated by the black arrow in (d), idealized as a point in the diagram). Repeated exposures of the same area reproduce the same pattern (g,h) but changing the position on the membrane shows a new configuration (i,j). The illuminated area in (g-j) has a diameter of ca. 150 nm, and ripples with a lateral extent down to the resolution limit of 15 nm are visible. (k) Colour-coded CBED intensities, representing a sample area of ca. 300 nm within the dashed line.

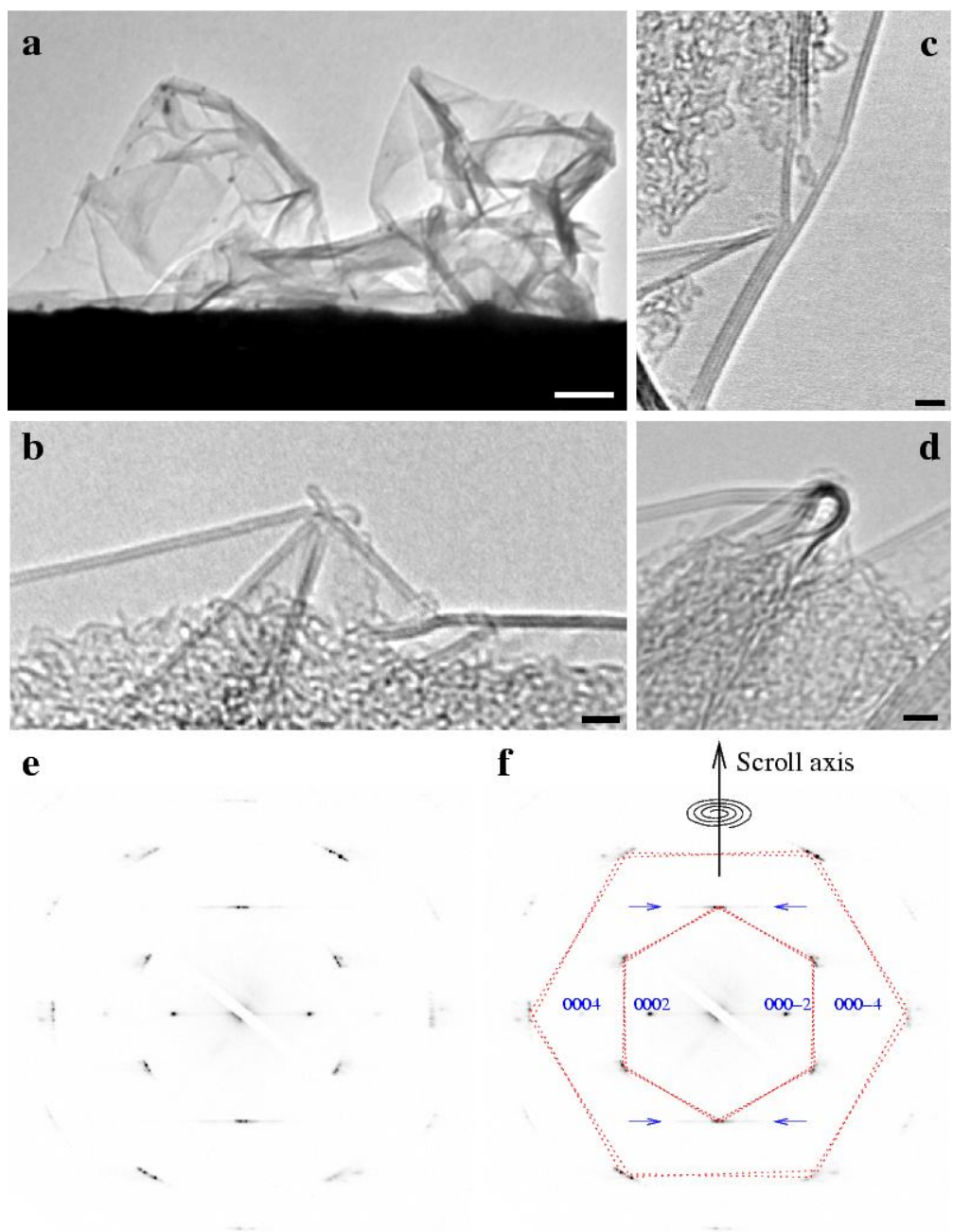

Fig. 8: Non-flat graphene configurations. (a) Overview of a folded flake that does not bridge the gap in the supporting grid but is attached only at one side. (b,c,d) Close-ups that resolve individual layers. (c) Transition from a single folded area (two layers) to a double fold (four layers). (d) Graphene sheet folded back onto itself and oriented parallel to the beam. The folding contains a hollow channel, similar in appearance and diameter to that typical for carbon nanotubes. We believe that similar nanoscrolls occur at the edges of flat areas such as shown in Fig. 4e,f. (e) Diffraction pattern from a scroll, and (f) the same pattern with features being assigned. The pattern is very similar to that of multi-walled carbon nanotubes. The dashed hexagons indicate two strong sets of peaks, showing that this scroll is rolled up almost along an armchair direction . The spacing and sharpness of (0002) type reflections shows that it is a tight scroll with an interlayer distance as in MWNTs (within experimental error). The streaks (indicated by arrows) show that the graphene sheet is indeed curved at the scrolls, rather than multiply folded. Scale bars (a) 50 nm, (b,c,d) 2 nm.